\documentclass[aps,amsmath,amssymb, reprint, prl]{revtex4-1}

\newcommand{\beqa}{\begin{eqnarray}}
\newcommand{\eeqa}{\end{eqnarray}}

\renewcommand{\k}{\kappa}
\newcommand{\half}{\frac{1}{2}}

\newcommand{\bea}{\begin{eqnarray}}
\newcommand{\eea}{\end{eqnarray}}

\newcommand{\CF}{{\mathcal F}}

\newcommand{\CN}{{\mathcal N}}

\def\e{\epsilon}

\def\k{\kappa}

\def\s{\sigma}

\def\G{\Gamma}

\def\CI{{\cal I}}

\newcommand{\be}{\begin{eqnarray}}
\newcommand{\ee}{\end{eqnarray}}
\newcommand{\nn}{\nonumber}
\newcommand{\bn}{\begin{enumerate}}
\newcommand{\en}{\end{enumerate}}

\def\tr{{\rm Tr}}

\def\ft{\mathfrak{t}}

\begin{document}

\title{Enhancement of Supersymmetry via Renormalization Group Flow\\ and the Superconformal Index}


\author{Kazunobu Maruyoshi}
\affiliation{Department of Physics, Imperial College London\\ Blackett Laboratory, Prince Concert Road, South Kensington, London, SW7 2AZ, UK}
\affiliation{
Faculty of Science and Technology, Seikei University\\ 3-3-1 Kichijoji-Kitamachi, Musashino-shi, Tokyo, 180-8633, Japan}
\email{maruyoshi@st.seikei.ac.jp}

\author{Jaewon Song}
\affiliation{Department of Physics, University of California, San Diego,  La Jolla, CA 92093, USA}
\email{jsong@physics.ucsd.edu}
\altaffiliation[Current address: ]{School of Physics, Korea Institute for Advanced Study, Seoul 02455, Korea}
\email{jsong@kias.re.kr}

\begin{abstract}
We find a four-dimensional $\CN=1$ gauge theory which flows to the minimal interacting $\CN=2$ superconformal field theory, the Argyres-Douglas theory, in the infrared up to the extra free chiral multiplets. The gauge theory is obtained from a certain $\CN=1$ preserving deformation of the $\CN=2$ $SU(2)$ gauge theory with four fundamental hypermultiplets. From this description, we compute the full superconformal index and find agreements with the known results in special limits.
\end{abstract}


\setcounter{tocdepth}{2}
\maketitle

\section{Introduction}
  Conformal field theory describes physics at a fixed point of a (quantum) gauge theory
  and therefore is one of the most important subjects in theoretical physics.
  It is well known that QCD with a particular amount of quarks flows to the infrared (IR) conformal fixed point.
  In four dimensions, however, it is difficult to probe conformal field theories in analytic level 
  because they are generically strongly coupled.  
  
  Symmetry is one of the most important properties for characterizing these theories.
  Apart from the theories with a spontaneously broken symmetry, it is often the case that 
  the symmetry of the IR conformal field theory is inherent in the ultraviolet (UV) theory.  
  However, sometimes the symmetry of the IR theory is emergent and not visible from the UV. 
  This is either due to a lack of proper formulation or the large quantum effects.  
  
  In this Letter, we present a theory realizing this phenomenon in a novel way:
  an $\CN=1$ supersymmetric gauge theory that flows to 
  the infrared fixed point governed by $\CN=2$ minimal superconformal field theory (SCFT), 
  namely, the Argyres-Douglas (AD) theory \cite{Argyres:1995jj}.
  Therefore, the infrared supersymmetry is enhanced from $\CN=1$ to $\CN=2$ in this model.
  
  Supersymmetry makes the theory rather tractable thanks to the techniques developed in recent decades, e.g., constraints from holomorphy \cite{Seiberg:1994bz}, localization \cite{Nekrasov:2002qd,Pestun:2007rz}, and so on.
  In particular, the superconformal index of a SCFT \cite{Romelsberger:2005eg,Kinney:2005ej}, or the partition function on $S^1 \times S^3$, 
  can be obtained from the localization technique.
  This quantity encodes the spectrum of the supersymmetry-protected sector. 
  When the SCFT is obtained as an IR fixed point of a Lagrangian theory, one can easily compute the index from the matter content in the UV. 
  
  The Argyres-Douglas theory obtained as the IR fixed point of our $\CN=1$ setup was originally found by considering the special locus in the Coulomb branch of $\CN=2$ supersymmetric gauge theory, where BPS states with mutually nonlocal electromagnetic charges become massless.
  Generalizations can be found in \cite{Argyres:1995xn, Eguchi:1996vu}, again as special loci in the Coulomb branches. This construction makes it impossible to write a Lagrangian for this theory.   Since the theory's discovery, not much has been known about it in the conformal phase, because of its lack of weakly coupled description.  
   
 Nevertheless, there are indications that the AD theory is the simplest or minimal $\CN = 2$ SCFT. As was shown in \cite{Beem:2013sza}, any $\CN = 2$ SCFTs have a protected sector described by the two-dimensional chiral algebra. For the AD theory and its generalizations, the corresponding chiral algebras are nonunitary minimal models \cite{Cordova:2015nma} or given by a simple coset \cite{Xie:2016evu}. In particular, the Argyres-Douglas theory that we find as the IR fixed point has the chiral algebra given by the simplest minimal model, namely, the Yang-Lee model. Moreover, the central charge $c$ takes the minimal value \cite{Liendo:2015ofa} among the interacting unitary four-dimensional $\CN=2$ SCFTs. 
    
  Our $\CN=1$ gauge theory description provides a new handle for studying aspects of this strongly interacting theory that has been mysterious. 
  The key ingredient in the analysis is the $a$-maximization \cite{Intriligator:2003jj} and its modification \cite{Kutasov:2003iy}. 
  This allows us to analyze the end point of the renormalization group (RG) flow, indicating that the IR theory is the minimal $\CN=2$ SCFT.  
  Furthermore, the gauge theory description enables us to compute various supersymmetric partition functions, in particular, the superconformal index. 
  The superconformal indices of the AD theory and its generalizations have been studied in \cite{Buican:2015ina, Cordova:2015nma, Buican:2015tda, Song:2015wta, Cecotti:2015lab}, but only in some particular limits. 
  Here, we compute the superconformal index in full generality.
  We find that the index computed in this way reproduces the previous results found in \cite{Cordova:2015nma,Song:2015wta}
  by taking the fugacity parameters to special values.

\section{The gauge theory} \label{sec:theory}

  Let us describe the gauge theory we study in this Letter. 
  First, consider $\CN=2$ supersymmetric $SU(2)$ gauge theory with $N_f=4$ fundamental hypermultiplets. 
  This theory preserves the $SO(8)$ global symmetry and has a moment map operator $\mu$ 
  which is the lowest component of the conserved current multiplet. 
  We then add a chiral multiplet $M$ transforming in the adjoint representation of $SO(8)$ and add the superpotential $W = \tr M \mu$. 
  Then we give a nilpotent vacuum expectation value (VEV) to $M$ given by $\rho(\s^+)$, 
  where $\rho$ is the embedding $\rho$: $\mathfrak{su}(2) \rightarrow \mathfrak{so}(8)$. 
  This is the type of deformation considered in \cite{Gadde:2013fma, Agarwal:2013uga,Agarwal:2014rua, Agarwal:2015vla}. 
  Depending on the choice of the embedding $\rho$, a different amount of $SO(8)$ flavor symmetry is broken. Here, we pick the principal embedding, which leaves no flavor symmetry. 
This will give masses to the fundamental quarks, and leave some components of $M$. See \cite{Maruyoshi:2016aim} for more details.

After integrating out massive components, we obtain the following gauge theory: 
there are two chiral multiplets transforming in the fundamental representation, one in the adjoint and four singlets coming from $M$. 
The charge assignment is as follows:
\begin{align}
\centering
\begin{array}{c|ccc}
  	& SU(2) & (J_+, J_-) & (R_0, \CF)  \\
	\hline
	q & \square & (1, 0) & (\half, \half) \\
	q' & \square & (1, -6) & (-\frac{5}{2}, \frac{7}{2}) \\
	\phi & {\bf adj} & (0, 2) & (1, -1) \\
	M_1 & 1 & (0, 4) & (2, -2)  \\
	M_3 & 1 & (0, 8) & (4, -4) \\
	M_5 & 1 & (0, 12) & (6, -6) \\
	M_3' & 1 & (0, 8) & (4, -4)
\end{array}
\end{align}
These are compatible with the superpotential 
\begin{align}
\begin{split}
 W = \phi q q &+ M_1 \phi^2 q q' + M_3 q q'   \\
  & + M_5 \phi q' q' + M_3' \phi^3 q' q',
\end{split}
\end{align}
where we omitted the gauge indices. 
The $U(1)_{J_\pm}$ are the nonanomalous $R$ symmetries coming from the Cartan parts of the $\CN=2$ $U(2)_R$. Therefore the superpotential should have charge $(J_+, J_-) = (2, 2)$. We also write the global symmetries as $U(1)_{R_0} \times U(1)_\CF$ for future convenience, given by $R_0 = \half (J_+ + J_-)$ and $\CF = \half (J_+ - J_-)$. If this theory flows to a SCFT in the IR, the superconformal $R$ symmetry will be given by a linear combination of the two $U(1)$s. Let us write
\be
 R_{{\rm IR}} = \frac{1+\e}{2} J_+ + \frac{1-\e}{2} J_- 
 = R_0 + \e \CF \ , 
 \label{RIR}
\ee
where the correct value of $\e$ at the superconformal point is determined via $a$-maximization \cite{Intriligator:2003jj}, as we will see shortly. 

\section{RG flow and $a$-maximization}
The central charge $a$ is given in terms of the 't Hooft anomaly coefficients of the IR superconformal $R$ symmetry as
  \be
  a &=&     \frac{3}{32} (3 \tr R^3 - \tr R)\ , \\
   c &=& \frac{1}{32}(9 \tr R^3 - 5 \tr R) \ .
  \ee
By substituting the expression \eqref{RIR}, the trial central charge $a(\epsilon)$ can be represented by the anomalies of $J_+$ and $J_-$.
For our theory, they are given by
\begin{align}
\begin{split}
	\begin{array}{c|c|c|c|c}
		J_+, J_+^3 & J_- & J_-^3 & J_+^2 J_- & J_+ J_-^2 \\
		\hline
		-4 & 18 & 1362 & 34 & -228 \\
	\end{array}
\end{split} \ , 
\end{align}
from which we get $a(\e) = - \frac{3}{32}  (807 \epsilon ^3-1746 \epsilon ^2+1231 \epsilon -284 )$. 
Upon $a$-maximization, we get $\e = \frac{1}{807} \left(582+\sqrt{7585}\right) \simeq 0.82911$ . 
This makes the Coulomb branch operator $\tr \phi^2$ [ which has $(J_+, J_-)=(0, 4)$ ] and $M_1$ violate the unitarity bound so that they become free along the RG flow and get decoupled. 

Let us redo the $a$-maximization after removing these chiral multiplets as in \cite{Kutasov:2003iy}. This gives the anomalies
\begin{align}
\begin{split}
	\begin{array}{c|c|c|c|c}
		J_+, J_+^3 & J_- & J_-^3 & J_+^2 J_- & J_+ J_-^2 \\
		\hline
		-2 & 12 & 1308 & 28 & -210 \\
	\end{array}
\end{split} \ , 
\end{align}
and 
this time we get $\e = \frac{1}{759} \left(558+\sqrt{8017}\right) \simeq 0.853146$.
With this value of $\e$, we find the $M_3$ and $M_3'$ operators violate the unitarity bound; thus, they get decoupled as well. 
Finally, after removing these operators, we get the anomalies
\begin{align}
\begin{split}
	\begin{array}{c|c|c|c|c}
		J_+, J_+^3 & J_- & J_-^3 & J_+^2 J_- & J_+ J_-^2 \\
		\hline
		0 & -2 & 622 & 14 & -112 \\
	\end{array}
\end{split} \ , 
\end{align}
and 
\begin{align}
 a(\e) &= - \frac{3}{32} \left(375 \epsilon ^3-810 \epsilon ^2+559 \epsilon -124\right) \ ,  \\
 c(\e) &= \frac{1}{32} \left(-1125 \epsilon ^3+2430 \epsilon ^2-1679 \epsilon +374\right) \ . 
\end{align}
By maximizing $a(\e)$, we obtain $\e = \frac{13}{15}$ and the central charges
\be
 a = \frac{43}{120} \ , \qquad c = \frac{11}{30} \ . 
\ee
These are exactly the same values as those of the AD theory \cite{Aharony:2007dj,Shapere:2008zf}. We also find that the operator $M_5$ has the conformal dimension $\Delta = \frac{6}{5}$, which is the value for the Coulomb branch operator of the AD theory. The value of the central charge $c=\frac{11}{30}$ is the minimal value of any interacting $\CN=2$ SCFT \cite{Liendo:2015ofa}. Therefore we claim that our gauge theory flows to the AD theory with four free chiral multiplets in the IR. In this sense, our construction gives a ``Lagrangian" description of the ``non-Lagrangian" AD theory. 

\section{Superconformal index} \label{sec:index}
As an application of our gauge theory description, let us compute the superconformal index or the partition function on $S^1 \times S^3$ of the AD theory. 

The superconformal index for the $\CN=1$ theory is defined as 
\be
 \CI_{\CN=1}(p, q, \xi) = \tr (-1)^F p^{j_1 + j_2 + \frac{R}{2}} q^{j_2 - j_1 + \frac{R}{2}} \xi^{\CF} , 
\ee
where $j_1$ and $j_2$ are the Cartan generators of the Lorentz group $SU(2)_1 \times SU(2)_2$, and $R$ and $\CF$ denote the generators of the $U(1)_R$ and the $U(1)_\CF$ symmetries respectively. 
While $R$ can be chosen to be any candidate $R$ charge, here we use $R_0$. 
After fixing the superconformal $R$ charge through the $a$-maximization, we redefine $\xi \to \xi (pq)^{\e/2}$ to get the proper index. 

Along the RG flow of our gauge theory, the operators $M_1, M_3, M_3'$ and $\tr \phi^2$ hit the unitarity bound and get decoupled. 
Therefore we should remove them from the index, similar to the prescription of \cite{Kutasov:2003iy}. 
%
 This gives us the integral
\begin{align}
 \CI_{UV} &= \k \frac{\G ((pq)^3 \xi^{-6})}{\G((pq)^1 \xi^{-2}) } \oint \frac{dz}{2\pi i z}  \\ 
 &~ \frac{ \G (z^\pm (pq)^{\frac{1}{4}} \xi^{\half}) \G (z^\pm (pq)^{-\frac{5}{4}} \xi^{\frac{7}{2}}) \G (z^{\pm 2, 0} (pq)^{\half} \xi^{-1})}{2 \G(z^{\pm 2})}, \nn 
\end{align}
where we used the abbreviation $f(z^\pm) \equiv f(z)f(z^{-1})$ and $f(z^{\pm 2, 0}) \equiv f(z^2)f(z^{-2})f(z^0)$. 
Here $\k = (p; p)(q; q)$ with $(z; q) = \prod_{n \ge 0} (1-z q^n)$ and $\G(z)$ is the elliptic gamma function
\be
 \G(z) \equiv \G(z; p, q) = \prod_{m, n \ge 0} \frac{1 - z^{-1} p^{m+1} q^{n+1}}{1- z p^m q^n} .
\ee
Each $\G$ factors in the numerator comes from each chiral multiplets and the factor $\k / \G(z^{\pm 2})$ is coming from the vector multiplet. The factor $\half$ is from the Weyl group of $SU(2)$. 
Note the term $\G((pq)^1 \xi^{-2})$ in the denominator: it is there to remove the contribution from the decoupled operator $\tr \phi^2$. We also removed the contributions from the singlets $M_1, M_3, M_3'$ in the integral. 
This should give us the integral that corresponds to the contribution of the interacting part of the theory only. 

Now, in order to obtain the index with the correct superconformal $R$ charge, we reparametrize $\xi \to \xi (pq)^{13/30}$. 
The $\CN=2$ superconformal index is defined as 
\be
 \CI_{\CN=2}(p, q, t) = \tr (-1)^F p^{j_1 + j_2 + r} q^{j_2 - j_1 + r} t^{I_3-r} , 
\ee
where $I_3$ and $r$ denotes the Cartan of $SU(2)_R$ symmetry and the generator of $U(1)_r$ symmetry respectively. 
We can map to the canonical $\CN=2$ fugacities by taking $\xi \to (t (pq)^{-\frac{2}{3}})^{\frac{1}{5}}$. 
In our convention, the $\CN=1$ $R$ charge, $R$, can be mapped to the $\CN=2$ $R$ charges as $R = \frac{2}{3}r + \frac{4}{3} I_3$. 
This gives the final expression
\begin{align}
\CI_{\CN=2} &= \k \frac{\G ((\frac{pq}{t})^{\frac{6}{5}})}{\G((\frac{pq}{t})^{\frac{2}{5}} ) } \oint_C \frac{dz}{2\pi i z} \\
&~\frac{\G (z^\pm (pq)^{\frac{2}{5}} t^{\frac{1}{10}}) \G (z^\pm (pq)^{-\frac{1}{5}} t^{\frac{7}{10}}) \G (z^{\pm 2, 0} (\frac{pq}{t})^{\frac{1}{5}} )}{2 \G (z^{\pm 2})}. \nn 
\end{align}
Here the contour of the integral $C$ (which is the unit circle around $z=0$) should include the pole at $z=(pq)^{-\frac{1}{5}} t^{\frac{7}{10}}$, but not at $z=(pq)^{\frac{1}{5}} t^{-\frac{7}{10}}$. See the related discussion in \cite{Agarwal:2014rua}. In practice, it is easier to consider reparametrization $p = \ft^3y, q=\ft^3/y, t = \ft^4/v$ and expand the integral in $\ft$. Let us write down the first few terms as a series expansion in $\ft$. We get
\bea \label{eq:ADidxInt}
 & &\CI_{\CN=2}(\ft, y, v)  = \tr (-1)^F \ft^{2(E+j_2)} y^{2j_1} v^{-(I_3+r)} 
 \nonumber \\
 & &~= 1+\ft^{\frac{12}{5}} v^{\frac{6}{5}}- \ft^{\frac{17}{5}} v^{\frac{1}{5}} \chi_2 (y) +\ft^{\frac{22}{5}}v^{-\frac{4}{5}} +\ft^{\frac{24}{5}} v^{\frac{12}{5}} \nn \\
& & ~~ + \ft^{\frac{27}{5}} v^{\frac{6}{5}} \chi_2 (y) - \ft^{\frac{29}{5}} v^{\frac{7}{5}} \chi_2 (y) -\ft^6 - \ft^{\frac{32}{5}} v^{\frac{1}{5}} \left(\chi_3 (y)+ \chi_1(y) \right) \nonumber \\
  & & ~~  + \ft^{\frac{34}{5}} v^{\frac{2}{5}}+\ft^7 v^{-1}  \chi_2 (y) +\ft^{\frac{36}{5}} v^{\frac{18}{5}}+ \ft^{\frac{37}{5}} v^{-\frac{4}{5}} \chi_2 (y) \nonumber  \\ 
  & &~~ +\ft^{\frac{39}{5}} v^{\frac{12}{5}} \chi_2 (y)  +\ft^8 v - \ft^{\frac{41}{5}}v^{\frac{13}{5}} \chi_2 (y)  \nonumber \\
 & &~~ + \ft^{\frac{42}{5}} v^{\frac{6}{5}} \left( \chi_3(y) - \chi_1 (y)\right) \nn \\
 & &~~ - \ft^{\frac{44}{5}} v^{\frac{7}{5}} \left( 2 \chi_3 (y) + \chi_1 (y) \right) - 2 \ft^9 \chi_2 (y) +O\left(\ft^{\frac{46}{5}}\right) , 
\eea
where $\chi_{2j_1+1} (y)$ is the character of the spin-$j$ representation of the $SU(2)_1$ rotation group. 

Let us now see some limits of the index.
We find that the Coulomb branch limit of the index ($\frac{pq}{t} = u$, $p, q, t \to 0$) is given by
\be
 \CI_C = \left( \frac{1-u^{\frac{2}{5}}}{1-u^{\frac{6}{5}}} \right) \oint \frac{dz}{2\pi i z} \frac{1 - z^{\pm 2}}{2( 1 - z^{\pm 2, 0} u^{\frac{1}{5}})} = \frac{1}{1-u^{\frac{6}{5}}} , \qquad
\ee
as expected, because the Coulomb branch is generated by a single operator with $\Delta = r = \frac{6}{5}$. 

Also, we find that the Macdonald limit $p \to 0$ indeed reproduces the leading order of the result given in \cite{Song:2015wta}
\be
 \CI_M = 1 + qt + q^2 t + q^3 t + \cdots \ . 
\ee
The absence of the $q^0 t^1$ term signals that there is no conserved current multiplet, which should contribute $\frac{t}{1-q}$ to the index. The term $\frac{qt}{1-q}$ comes from the stress tensor multiplet. Note that there is no term of the form $q^2 t^2$, which means that a short multiplet that generally appears in the OPE $T \times T$ is absent in this theory. This is precisely the condition to saturate the bound $c \ge\frac{11}{30}$ derived in \cite{Liendo:2015ofa}.  

Furthermore, we reproduce the Schur limit $t \to q$ of the index predicted in \cite{Cordova:2015nma} in the leading order as well. It would be nice to prove that the integral formula \eqref{eq:ADidxInt} indeed reproduces the closed-form formula for the Macdonald index given in \cite{Song:2015wta} or the Schur index given by the vacuum character of the Yang-Lee model
\be
 \chi^{c=-\frac{22}{5}}_0 (q) =  \sum_{n \ge 0} \frac{q^{n^2 + n}}{(q)_n} = \frac{1}{(q^2; q^5)(q^3; q^5)}  , 
\ee
where $(q)_n \equiv \prod_{m=1}^{n} (1-q^m)$ and the second equality is the Rogers-Ramanujan identity. 

\section{Discussion}
In this Letter, we have found an $\CN=1$ gauge theory obtained by the deformation of the $\CN=2$ $SU(2)$ gauge theory with four fundamental hypermultiplets, which realizes the minimal $\CN=2$ SCFT at the end point of the RG flow. We find that our $\CN=1$ theory exhibits an emergent extended $\CN=2$ supersymmetry. This theory provides a handle for investigating various aspects of the non-Lagrangian AD theory which was previously inaccessible. As an application, we computed the full superconformal index. 
 Let us make a few comments.

The deformed theory we study (after integrating out the massive modes by using the Higgs mechanism) has the matter content similar to the $\CN=2$ $SU(2)$ $N_f=1$ gauge theory, except for the four extra singlets. This is closely related to the setup used to obtain the AD theory in \cite{Argyres:1995xn}, where they set the mass parameters and move into a particular point in the Coulomb branch. It appears that our superpotential and extra chiral multiplets have the effect of setting the relevant parameters to be the special value required to be at the point for the AD theory. It is interesting to ask whether there is a generic way of engineering such a flow, which may give us a way to obtain Lagrangian descriptions for the other SCFTs as well. 

It is widely believed that every $\CN=2$ SCFT (except for a free hypermultiplet) has a Coulomb branch, and the AD theory is no exception. For our gauge theory, it is the singlet operator $M_5$ that ends up being the chiral operator of the IR theory parametrizing the Coulomb branch. However, it is unclear from the gauge theory perspective why giving an expectation value to this operator should cause the theory to be in the Coulomb phase. It would be interesting to understand how the Coulomb phase appears in the IR. 

There is a similar result realizing the $\CN=2$ $E_6$ SCFT \cite{Minahan:1996fg} as the end point of the RG flow of an $\CN=1$ gauge theory \cite{Gadde:2015xta}. This is somewhat similar to our result, but the way each model works is quite different. In our case, some of the operators decouple along the RG flow due to the accidental global symmetry. Moreover, the Coulomb branch appears at the end of the RG flow is not visible from the high energy. 

Finally, we point out that the partition function of the AD theory on other manifolds can be computed by using our gauge theory description. It would be interesting to further develop this direction. 

\vspace{6pt}

\begin{acknowledgments}
We would like to thank Prarit Agarwal, Philip Argyres, Ken Intriligator and Yuji Tachikawa for helpful discussions and comments. 
We thank Prarit Agarwal for sharing his note on the accidental symmetries and superconformal index in adjoint SQCD.
The work of K.~M.~is supported by the EPSRC Programme Grant No.~EP/K034456/1, ``New Geometric Structures from String Theory.''
The work of J.~S.~is supported in part by the U.~S.~Department of Energy under UCSD's Contract No.~de-sc0009919 and also by the Hwa-Ahm foundation.
\end{acknowledgments}

\bibliography{ADLag}

\end{document}